# Eliminating ground-state dipole moments in quantum optics via canonical transformation

*Gediminas Juzeliunas*[*] *Luciana Dávila Romero*[†] *and David L. Andrews*[†]

[*]*Institute of Theoretical Physics and Astronomy, Vilnius University, A. Goštauto 12, 2600 Vilnius, Lithuania*

[†]*School of Chemical Sciences and Pharmacy, University of East Anglia, Norwich NR4 7TJ, United Kingdom*

## Abstract

By means of a canonical transformation it is shown how it is possible to recast the equations for molecular nonlinear optics to completely eliminate ground-state static dipole coupling terms. Such dipoles can certainly play a highly important role in nonlinear optical response – but equations derived by standard methods, in which these dipoles emerge only as special cases of transition moments, prove unnecessarily complex. It has been shown that the elimination of ground-state static dipoles in favor of dipole shifts results in a considerable simplification in form of the nonlinear optical susceptibilities. In a fully quantum theoretical treatment the validity of such a procedure has previously been verified using an expedient algorithm, whose defense was afforded only by a highly intricate proof. In this paper it is shown how a canonical transformation method entirely circumvents such an approach; it also affords new insights into the formulation of quantum field interactions.

## 1. Introduction

In recent years it has become increasingly evident that permanent (static) electric dipoles play a highly important role in the nonlinear optical response of molecular systems [1, 2]. Most molecules are intrinsically polar by nature, and calculation of their optical susceptibilities with regard only to transition dipole moments can produce results that are significantly in error [3, 4]. In particular, many of the 'push-pull' systems favored for their high degree of optical nonlinearity are specifically those where permanent dipole effects are the largest, through their designed juxtaposition of strongly electron-donating and electron-withdrawing functional groups [5-7].

Whilst a number of groups have developed the theory to elicit permanent dipole contributions to nonlinear optical response, the framework for most of this work has been semiclassical [8, 9]. In such a context, where the molecular system is treated with quantum mechanical rigour but the radiation is treated as a classical oscillatory electric field, it has been demonstrated that a transformed 'fluctuation dipole' Hamiltonian properly describes the optical interactions of the dynamical system, and affords considerable calculational simplification [4]. However the semiclassical treatment fails to take account of associated modifications to electromagnetic field interactions – features that only emerge in a quantum electrodynamical treatment. Indeed it has been remarked that quantum electrodynamics affords the only completely rigorous basis for descriptions of multipolar behavior [10, 11].

When both the material and radiation parts of the system are developed in fully quantised form, the transparency and correctness of deploying a transformed interaction Hamiltonian is potentially obscured [12]. Both the conventional and transformed operators prove to lead to identical results even at high orders of optical nonlinearity [13], yet the equivalence of their predictions as a general principle has until now been established only through engagement in proofs of considerable intricacy [14]. It is our purpose in this paper to rectify this anomaly by demonstrating the implementation of a suitable canonical transformation on the multipolar form of the quantum optical Hamiltonian. Elucidating the quantum electrodynamical treatment in this way throws new light on a number of issues skirted over in the semiclassical treatment, and it offers clear scope for extension to higher orders of multipole interaction.

Employment of the canonical transformation effects a considerable simplification of the analysis of nonlinear optical processes involving permanent dipole moments. For instance, in the standard formulation, second harmonic generation is represented by $3 \times 2^2 = 12$ different state sequences when a two-level molecular model is used. Each of these entails a product of three 'transition' dipoles (one or more of which may be permanent), divided by a product of two energy factors. However, the new Hamiltonian will involve only three terms of simpler structure containing no explicit contributions due to the ground state dipole moment, only dipole shifts (i.e. differences between excited state and ground state moments) – as we shall see in Sec.4.2.

The paper is organized as follows. In the next section the radiation-matter Hamiltonian is introduced – taking the leading electric dipole terms from the multipolar formulation of Quantum Electrodynamics (QED). In Sec. 3.1 a canonical transformation is carried out to eliminate the coupling between the quantised radiation field and permanent dipoles of the molecules in ground electronic states. In Sec.3.2 the Hamiltonian is re-expressed in the new representation, followed by an application to the study of nonlinear optical processes in Sec.4. The example of second harmonic generation (SHG) studied in detail in Sec. 4.2 illustrates how the new formalism facilitates elimination of the contributions due to the ground state dipole in appropriately time-ordered quantum channels for the molecule-radiation processes. Extensions beyond the dipole approximation are considered in Sec.5, followed by concluding remarks in Sec.6.

## 2. The multipolar QED Hamiltonian

We begin with the Hamiltonian in multipolar form, for a system of molecules labeled $X$, interacting with a quantised radiation field [15]:

$$H = H_{rad} + \sum_X H_{mol}(X) + H_{int}. \qquad (1)$$

where $H_{rad}$ is the Hamiltonian for the free radiation field, each $H_{mol}(X)$ is a Schrödinger operator for an isolated molecule, and $H_{int}$ is a term representing a fully retarded coupling between the quantised radiation field and the molecular subsystem.

The first of these operators, $H_{rad}$, is expressed as follows in terms of the transverse electric displacement field operator $\mathbf{d}^\perp(\mathbf{r})$ and magnetic field operator $\mathbf{b}(\mathbf{r})$:

$$H_{rad} = \tfrac{1}{2}\int\left\{\varepsilon_0^{-1}\left[\mathbf{d}^\perp(\mathbf{r})\right]^2 + \mu_0^{-1}\left[\mathbf{b}^\perp(\mathbf{r})\right]^2\right\}d^3\mathbf{r}, \qquad (2)$$

with

$$\mathbf{d}^\perp(\mathbf{r}) = \varepsilon_0\,\mathbf{e}^\perp(\mathbf{r}) + \mathbf{p}^\perp(\mathbf{r}). \qquad (3)$$

In equation (3), $\mathbf{e}^\perp(\mathbf{r})$ is the transverse part of the electric field, in the electric-dipole approximation, and $\mathbf{p}(\mathbf{r})$ is the polarization field of molecular origin, given by

$$\mathbf{p}(\mathbf{r}) = \sum_X \boldsymbol{\mu}(X)\,\delta(\mathbf{r}-\mathbf{R}_X), \qquad (4)$$

$\mathbf{p}^\perp(\mathbf{r})$ being its transverse part [15]. Again, $\boldsymbol{\mu}(X)$ is the dipole operator of the molecule $X$ positioned at $\mathbf{R}_X$, and the summations are taken over all the molecules of the system. Next, the Hamiltonian for molecule $X$ is explicitly given by:

$$H_{mol}(X) = \frac{1}{2m}\sum_a p_{X,a}^2 + \frac{e^2}{4\pi\varepsilon_0}\sum_{a,b}\frac{1}{|\mathbf{q}_{X,a}-\mathbf{q}_{X,b}|}, \qquad (5)$$

where $\mathbf{p}_{X,a}$ and $\mathbf{q}_{X,a}$ are, respectively, operators for the momentum and position of electron $a$. Finally the operator $H_{int}$, which describes the coupling between the molecular sub-system and the quantised radiation field, is expressible as:

$$H_{int} = -\int \varepsilon_0^{-1}\,\mathbf{d}^\perp(\mathbf{r})\,\mathbf{p}(\mathbf{r})\,d^3\mathbf{r} = -\sum_X \varepsilon_0^{-1}\,\mathbf{d}^\perp(\mathbf{R}_X)\cdot\boldsymbol{\mu}(X). \qquad (6)$$

Although the field interaction is here cast in terms of the electric dipole approximation, our analysis can be extended quite straightforwardly to incorporate higher multipole terms – this will be discussed in Sec.5. Lastly, the dipole operator $\boldsymbol{\mu}(X)$ can be cast in matrix form in terms of molecular dipole moments;

$$\boldsymbol{\mu}(X) = \sum_{j,l} \left| j^{(X)} \right\rangle \boldsymbol{\mu}_{jl}^{(X)} \left\langle l^{(X)} \right| , \qquad (7)$$

where $\left| j^{(X)} \right\rangle$ are the eigenvectors of the molecular Hamiltonian $H_{mol}(X)$. In the multipolar QED formulation employed [15-17], the Hamiltonian of the system does not contain any instantaneous dipole-dipole type interactions, all intermolecular coupling being mediated via the transverse radiation field. Specifically, each molecule is coupled to the displacement field at the molecular site, as is evident in Eq.(6).

## 3. Transformation to another representation

### 3.1 Canonical transformation

To eliminate the permanent ground-state dipole $\boldsymbol{\mu}_{gg}^{(X)}$, we shall apply the following canonical transformation, which recasts operators but effects no change in any system observables:

$$u = \exp(iS); \qquad S = -(1/\hbar) \int \mathbf{a}^{\perp}(\mathbf{r}) \, \mathbf{p}_g(\mathbf{r}) \, d^3\mathbf{r}, \qquad (8)$$

with

$$\mathbf{p}_g(\mathbf{r}) = \sum_X \boldsymbol{\mu}_{gg}^{(X)} d(\mathbf{r} - \mathbf{R}_X) ; \qquad (9)$$

the latter signifies the polarization field produced by an assembly of ground-state molecules containing static dipoles, $\boldsymbol{\mu}_{gg}^{(X)}$. The transformation does not alter the vector potential $\mathbf{a}^\perp(\mathbf{r})$, since the generator $S$ commutes with it. Furthermore, since the dipole moment $\boldsymbol{\mu}_{gg}^{(X)}$ is a *c*-number characterized by a real value (i.e. it is not an operator), the transformation does not modify either the electron momenta $\mathbf{p}_{X,a}$ or coordinates $\mathbf{q}_{X,a}$ featured in the molecular Hamiltonian $H_{mol}(X)$. The only canonical variable affected by the transformation is the electric displacement operator, and from the commutator relation $\left[a_i^\perp(\mathbf{r}), d_j^\perp(\mathbf{r}')\right] = -i\hbar d_{ij}^\perp(\mathbf{r}-\mathbf{r}')$ we have:

$$\tilde{\mathbf{d}}^\perp(\mathbf{r}) \equiv e^{iS}\mathbf{d}^\perp(\mathbf{r})e^{-iS} = \mathbf{d}^\perp(\mathbf{r}) - \mathbf{p}_g^{\perp}(\mathbf{r}), \qquad (10)$$

i.e. the transformation effects a subtraction from the full displacement field $\mathbf{d}^\perp(\mathbf{r})$ of the transverse part of the polarization field due to the ground-state dipoles, $\boldsymbol{\mu}_{gg}^{(X)}$. Alternatively, one can write:

$$\tilde{\mathbf{d}}^\perp(\mathbf{r}) = e_0 \mathbf{e}^\perp(\mathbf{r}) + \tilde{\mathbf{p}}^\perp(\mathbf{r}), \qquad (11)$$

where

$$\tilde{\mathbf{p}}(\mathbf{r}) \equiv \mathbf{p}(\mathbf{r}) - \mathbf{p}_g(\mathbf{r}) = \sum_X \tilde{\boldsymbol{\mu}}(X)\, d(\mathbf{r}-\mathbf{R}_X) \qquad (12)$$

is the polarization field *excluding* the ground-state dipoles, and

$$\tilde{\boldsymbol{\mu}}(X) \equiv \boldsymbol{\mu}(X) - \boldsymbol{\mu}_{gg}^{(X)} = \sum_{j,l}\left| j^{(X)} \right\rangle \tilde{\boldsymbol{\mu}}_{lj}^{(X)} \left\langle l^{(X)} \right| \qquad (13)$$

is the corresponding dipole operator, with $\tilde{\boldsymbol{\mu}}_{lj}^{(X)} = \boldsymbol{\mu}_{lj}^{(X)} - \boldsymbol{\mu}_{gg}^{(X)} d_{lj}$.

### 3.2 Hamiltonian in the new representation

Substituting $\mathbf{d}^\perp(\mathbf{r}) = \tilde{\mathbf{d}}^\perp(\mathbf{r}) + \mathbf{p}_g^\perp(\mathbf{r})$ into equations (2) and (6) the Hamiltonian of the system, equation (1) can be re-expressed as:

$$H = \tilde{H}_{rad} + \sum_X \tilde{H}_{mol}(X) + \tilde{H}_{int}, \qquad (14)$$

where

$$\tilde{H}_{rad} = 2^{-1} \int \left\{ \varepsilon_0^{-1} \left[\tilde{\mathbf{d}}^\perp(\mathbf{r})\right]^2 + \mu_0^{-1} \left[\mathbf{b}^\perp(\mathbf{r})\right]^2 \right\} d^3\mathbf{r}, \qquad (15)$$

and

$$\tilde{H}_{int} = -\int \varepsilon_0^{-1} \tilde{\mathbf{d}}^\perp(\mathbf{r}) \tilde{\mathbf{p}}(\mathbf{r}) d^3\mathbf{r} = -\sum_X \varepsilon_0^{-1} \tilde{\mathbf{d}}^\perp(\mathbf{R}_X) \cdot \tilde{\boldsymbol{\mu}}^{(X)} \qquad (16)$$

are the radiative and interaction Hamiltonian in the new representation. In this way, the radiation-molecule coupling is now represented in terms of the dipole moment operator $\tilde{\boldsymbol{\mu}}^{(X)} = \boldsymbol{\mu}^{(X)} - \boldsymbol{\mu}_{gg}^{(X)}$ which excludes a contribution associated with the permanent ground-state dipole moment. The new molecular Hamiltonian can be written as:

$$\tilde{H}_{mol}(X) = H_{mol}(X) + V_{mol}(X), \qquad (17)$$

where the extra term reads:

$$V_{mol}(X) = -\frac{1}{2} \sum_{X' \neq X} \frac{\boldsymbol{\mu}_{gg}^{(X')} \boldsymbol{\mu}_{gg}^{(X)} - 3\left(\boldsymbol{\mu}_{gg}^{(X')} \cdot \hat{\mathbf{R}}_{XX'}\right)\left(\hat{\mathbf{R}}_{XX'} \cdot \boldsymbol{\mu}_{gg}^{(X)}\right)}{4\pi\varepsilon_0 R_{XX'}^3} - \sum_{X' \neq X} \frac{\boldsymbol{\mu}_{gg}^{(X')} \tilde{\boldsymbol{\mu}}^{(X)} - 3\left(\boldsymbol{\mu}_{gg}^{(X')} \cdot \hat{\mathbf{R}}_{XX'}\right)\left(\hat{\mathbf{R}}_{XX'} \cdot \tilde{\boldsymbol{\mu}}^{(X)}\right)}{4\pi\varepsilon_0 R_{XX'}^3}$$

$$(18)$$

with $\mathbf{R}_{XX'} = \mathbf{R}_X - \mathbf{R}_{X'}$. The condition $X' \neq X$ ensures omission of molecular self-interaction terms (otherwise known as *contact interaction* terms).

The first term in (18) can be identified as the direct (Coulomb) interaction between the ground state dipole of a particular molecule $X$ and those of all other species $X'$, the factor of ½ preventing double counting when we sum over all $X$ centers – see equations (14) and (17). The second term in (18) can be identified as a contribution due to Coulomb coupling between the transition dipoles of a selected molecule $X$ and the ground state dipoles of surrounding species $X'$. Both types of molecule-molecule coupling are instantaneous, because they are associated with at least one permanent dipole.

The emergence of the instantaneous intermolecular coupling is a direct consequence of eliminating the permanent ground-state dipoles from the field-mediated molecule-molecule coupling in the reformulated multipolar QED representation. The additional interaction $V_{mol}(X)$ in Eq.(17) effects a modification to the eigenstates and corresponding energies of molecule $X$ due to surrounding molecules with permanent dipole moments. For nonlinear and other quantum optical processes in dilute molecular systems, where molecular centers are involved only singly, the extra term $V_{mol}(X)$ does not intervene and molecular states and energy levels remain essentially unchanged.

## 4. Nonlinear optical processes

### 4.1 Introduction

With the Hamiltonian now re-expressed in the new representation, it is possible to apply it to the study of nonlinear optical processes. The example of second harmonic generation (SHG) to be studied in detail illustrates how this facilitates the elimination of ground state dipole terms in a rigorous QED treatment. This circumvents the highly intricate proof that previously afforded its only justification in QED – and even then, only for two-level systems [13]. It is important to emphasize that the transformed displacement field operator and the radiation Hamiltonian can be cast in the usual form [15]:

$$\tilde{\mathbf{d}}^{\perp}(\mathbf{R}) = i \sum_{\mathbf{k},l} \left( \frac{\hbar c k e_0}{2V} \right)^{1/2} \mathbf{e}^{(l)}(\mathbf{k}) \left\{ a^{(l)}(\mathbf{k}) e^{i\mathbf{k}.\mathbf{R}} - a^{(l)+}(\mathbf{k}) e^{-i\mathbf{k}.\mathbf{R}} \right\} \tag{19}$$

and

$$\tilde{H}_{rad} = \sum_{\mathbf{k},l} \left\{ a^{(l)+}(\mathbf{k}) a^{(l)}(\mathbf{k}) + 1/2 \right\} \hbar c \mathbf{k} , \tag{20}$$

where, in each expression, a sum is taken over radiation modes characterized by wave-vector $\mathbf{k}$ and unit polarization vector $\mathbf{e}^{(l)}(\mathbf{k})$ (with $l$ denoting one of two polarization states); $a^{(l)+}(\mathbf{k})$ and $a^{(l)}(\mathbf{k})$ are the corresponding operators for creation and annihilation of a photon, and $V$ is an arbitrarily large quantization volume. Note that the photon states are somewhat different to those emerging in the original multipolar QED, because the canonical operator $\tilde{\mathbf{d}}^{\perp}(\mathbf{R})$ differs from the original displacement operator $\mathbf{d}^{\perp}(\mathbf{r})$ by the amount $-\mathbf{p}_g^{\perp}(\mathbf{r})$.

From expression (19), the application of perturbation theory for the study of optical processes is straightforward and follows similar lines to those employed when the multipolar formalism is used [15]. For simplicity and clarity we take the dilute gas approximation, dispensing with the refractive effects in the mode-expansion of the displacement field operator (19). Nonetheless, our method is amenable to the incorporation of such effects; the displacement field operator is then expanded in terms of photons fully dressed by the molecular medium (i.e. polaritons) [18-21] rather than the 'bare' photons.

**4.2 Second harmonic generation**

Second harmonic generation serves to illustrate use of the new Hamiltonian $H$, equation (14). This well-known optical process can be described as fundamentally involving the annihilation of two photons of a certain frequency $w$ and the creation of one photon of double the frequency, $2w$. The molecular centers are initially in their ground state: $|i^{(X)}\rangle \equiv |g^{(X)}\rangle$. Since SHG is an elastic process, the final molecular states are identical to the initial ones. The initial state of the radiation field, $|n(\mathbf{k},\mathbf{l})\rangle$, contains $n$ photon in a particular electromagnetic mode $(\mathbf{k},\mathbf{l})$, while the final radiative state is $|n-2(\mathbf{k},\mathbf{l});1(\mathbf{k}',\mathbf{l})\rangle$, with $|\mathbf{k}'|=2|\mathbf{k}|$. The optical process can be represented by three time-ordered diagrams [12, 13] (see also figure 1). Each of these time-ordered diagrams in turn contributes various terms depending on the number of molecular states involved in the optical process. In the conventional formulation, the number of terms can be fairly large. For example, if the molecules are represent by a two-level model, the transition matrix will possess $3 \times 2^2 = 12$ contributions - each a product of three 'transition' dipoles (one or more of which may be permanent), divided by a product of two energy factors. However, the new Hamiltonian involves only three simpler terms containing only contributions directly associated with dipole shifts. In the more general case, lifting the two-level approximation, the transition amplitude of one centre reads in the new representation:

$$S_X \sim \langle (n-2)(\mathbf{k},\mathbf{l}),1(\mathbf{k}',\mathbf{l}); X,i | \left\{ \tilde{H}_{int} \frac{1}{(\tilde{E}_0 - \tilde{H}_0)} \tilde{H}_{int} \frac{1}{(\tilde{E}_0 - \tilde{H}_0)} \tilde{H}_{int} \right\} | n(\mathbf{k},\mathbf{l}); X,i \rangle$$

(21)

where $\tilde{E}_0 = E_i + n\hbar w$ is the energy of the initial state of the system, and the new unperturbed Hamiltonian $\tilde{H}_0$ is given by $\tilde{H}_{rad} + \sum_X H_{mol}(X)$. Again, since the molecular system is considered to be dilute, we have neglected the medium-induced term $V_{mol}(X)$ in the molecular Hamiltonian $\tilde{H}_{mol}(X) = H_{mol}(X) + V_{mol}(X)$.

When equations (16) and (19) are used the transition matrix (21) can be expressed as:

$$S_X = -i \left( \frac{\hbar c}{2\mathbf{e}_o V} \right)^{3/2} (k^2 k')^{1/2} \{n(n-1)\}^{1/2} \vec{e}_i' e_j e_k \mathbf{b}_{ijk},$$ (22)

where $\mathbf{b}_{ijk}$ is the hyperpolarisability tensor given by

$$\mathbf{b}_{ijk} = \sum_{r,s} \frac{\tilde{m}_i^{0r} \tilde{m}_j^{rs} \tilde{m}_k^{s0}}{(\tilde{E}_{0r} + 2\hbar w)(\tilde{E}_{0s} + \hbar w)} + \frac{\tilde{m}_j^{0r} \tilde{m}_i^{rs} \tilde{m}_k^{s0}}{(\tilde{E}_{0r} - \hbar w)(\tilde{E}_{0s} + \hbar w)} + \frac{\tilde{m}_j^{0r} \tilde{m}_k^{rs} \tilde{m}_i^{s0}}{(\tilde{E}_{0r} - \hbar w)(\tilde{E}_{0s} - 2\hbar w)}.$$ (23)

with ω = ck. In passing we note that in any application the index-symmetrised form, $b_{i(jk)} \equiv \tfrac{1}{2}\left(b_{ijk} + b_{ikj}\right)$, would necessarily be invoked because of the corresponding symmetry in the radiation tensor – with which it is eventually contracted to give a result for the signal. The result given by equation (23) is correct for molecular cases far from resonance. If near-resonant terms are considered then phenomenological damping factors are introduced. By adopting the convention of a constant sign for the damping [22], as recently confirmed [23], the result coincides exactly with earlier work, and now without any need to assume that the linewidth is small compared to the harmonic frequency.

## 5. Extension beyond the dipole approximation

The above analysis can be extended beyond the dipole approximation. For this purpose one needs to add non-dipole contributions to the polarization field $\mathbf{p}^{\perp}(\mathbf{r})$ featured in the displacement field given by Eq.(3). Furthermore, one needs to include the non-dipole terms in the multipolar Hamiltonian [15]. Subsequently, one can transform such a full multipolar Hamiltonian *via* the canonical transformation of the form of equation (8) that excludes the polarization field not only due to static dipoles $\boldsymbol{\mu}_{gg}^{(X)}$ but also the corresponding higher order electric and magnetic multipoles. This will lead to the Hamiltonian of the same form as equation (14), in which the interaction Hamiltonian $\tilde{H}_{\mathrm{int}}$ now accommodates the full multipolar expansion of the radiation-matter coupling (as presented explicitly in ref.15), subject to replacement not only of the dipole-operator $\boldsymbol{\mu}^{(X)}$ by $\tilde{\boldsymbol{\mu}}^{(X)}$, but also with corresponding modifications to the higher-order moments. Furthermore, there will be an additional contribution in the operator $V_{mol}(X)$ due to the coupling between the multipoles of a specific molecule $X$ and the static multipoles of the remaining species. In this way electrostatic interactions due to the permanent multipoles of the ground state are included in the Hamiltonian for the electrostatic intermolecular coupling. As such, the method extends the applicability of the calculational algorithm [13], previously directed only at electric dipole interactions, offering further scope for simplifying the formulation of theory for optical processes involving permanent multipoles.

## 6. Concluding remarks

A canonical transformation has been introduced to completely eliminate ground state dipole coupling terms in the multipolar formulation of quantum electrodynamics. The transformation does not alter the vector potential $\mathbf{a}^\perp(\mathbf{r})$, yet it effects a subtraction from the full displacement field $\mathbf{d}^\perp(\mathbf{r})$ of the transverse part of the polarization field due to the ground-state dipoles, $\boldsymbol{\mu}_{gg}^{(X)}$. The radiation-molecule coupling is then represented in terms of the dipole moment operator $\tilde{\boldsymbol{\mu}}^{(X)} = \boldsymbol{\mu}^{(X)} - \boldsymbol{\mu}_{gg}^{(X)}$ excluding a contribution due to the permanent ground-state dipole moment. An additional instantaneous intermolecular interaction appears in the transformed Hamiltonian, representing changes in the molecular eigenstates and corresponding energies due to a surrounding polar medium. The emergence of instantaneous intermolecular coupling is a direct consequence of eliminating, in the field-mediated molecule-molecule coupling, the permanent ground-state dipoles in favor of dipole shifts.

The new canonical transformation method concisely circumvents a highly intricate proof that previously afforded the only rigorous justification. Moreover the quantum electrodynamical treatment elucidates a number of issues skirted over in the semiclassical treatment. The QED method is directly amenable to the inclusion of higher rank multipole moments; furthermore it extends previous work in permitting application to molecules with an arbitrary number of molecular states. This allows the proper representation of non-resonant optical processes, and is fully consistent with the constant sign convention for phenomenological damping. The method leads directly to susceptibility results cast in the simplest possible form, as has been illustrated with the example of second harmonic generation.

## 7. Acknowledgments

We are grateful to Guy Woolley and Mohamed Babiker for helpful comments. GJ acknowledges the Royal Society for funding his visit to the United Kingdom. DLA and LCDR acknowledge financial support from the Engineering and Physical Sciences Research Council.

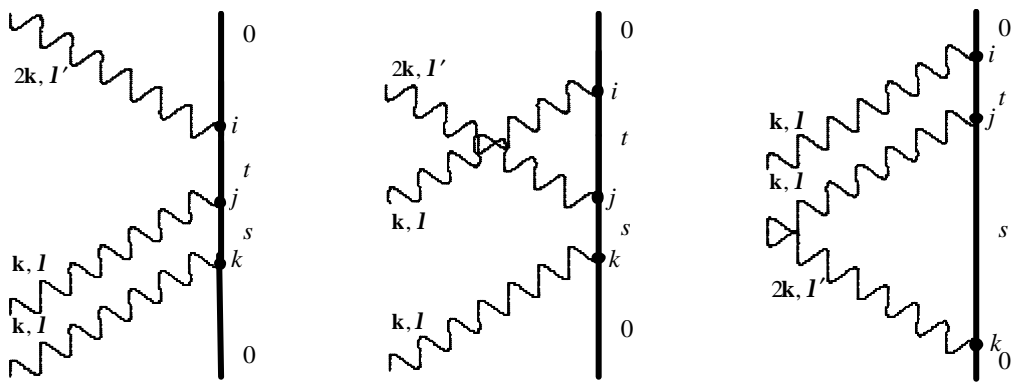

**Figure 1:** The three time ordered diagrams for second harmonic generation